\documentclass[twocolumn,aps,prl,superscriptaddress,showpacs,byrevtex]{revtex4-2}
\usepackage{epsf,graphicx,amssymb}
\usepackage[usenames]{color}
\usepackage{natbib}
\usepackage{float,ulem}
\usepackage{gensymb}
\usepackage{amsmath,amssymb}
\usepackage{color}
\usepackage{empheq}
\usepackage{url}

\begin{document}

\title{Structural Order Drives Diffusion in a Granular Packing}

\author{David Luce}
\affiliation{LEMTA, Université de Lorraine, CNRS, 2, Avenue de la Forêt de Haye, Vandœuvre-lès-Nancy, 54504, France} 
\affiliation{GRASP, Institute of Physics B5a, Université de Liège, 4000 Liège, Belgium}
\author{Adrien Gans} 
\affiliation{LEMTA, Université de Lorraine, CNRS, 2, Avenue de la Forêt de Haye, Vandœuvre-lès-Nancy, 54504, France}
\author{Sébastien Kiesgen De Richter}
\affiliation{LEMTA, Université de Lorraine, CNRS, 2, Avenue de la Forêt de Haye, Vandœuvre-lès-Nancy, 54504, France}
\affiliation{Institut Universitaire de France (IUF)} 
\author{Nicolas Vandewalle} 
\affiliation{GRASP, Institute of Physics B5a, Université de Liège, 4000 Liège, Belgium}

\begin{abstract}
We investigate how structural ordering, i.e. crystallization, affects the flow of bidisperse granular materials in a quasi-two-dimensional silo. By systematically varying the mass fraction of two particle sizes, we finely tune the degree of local order. Using high-speed imaging and kinematic modeling, we show that crystallization significantly enhances the diffusion length $b$, a key parameter controlling the velocity profiles within the flowing medium. We reveal a strong correlation between $b$ and the hexatic order parameter $\psi_6$, highlighting the role of local structural organization in governing macroscopic flow behavior. Furthermore, we demonstrate that pressure gradients within the silo promote the stabilization of orientational order even in the absence of crystallization, thus intrinsically increasing $b$ with height. These findings establish a direct link between microstructural order, pressure, and transport properties in granular silo flows.
\end{abstract}

\maketitle

\begin{figure}[h!]
\includegraphics[width=0.49\textwidth]{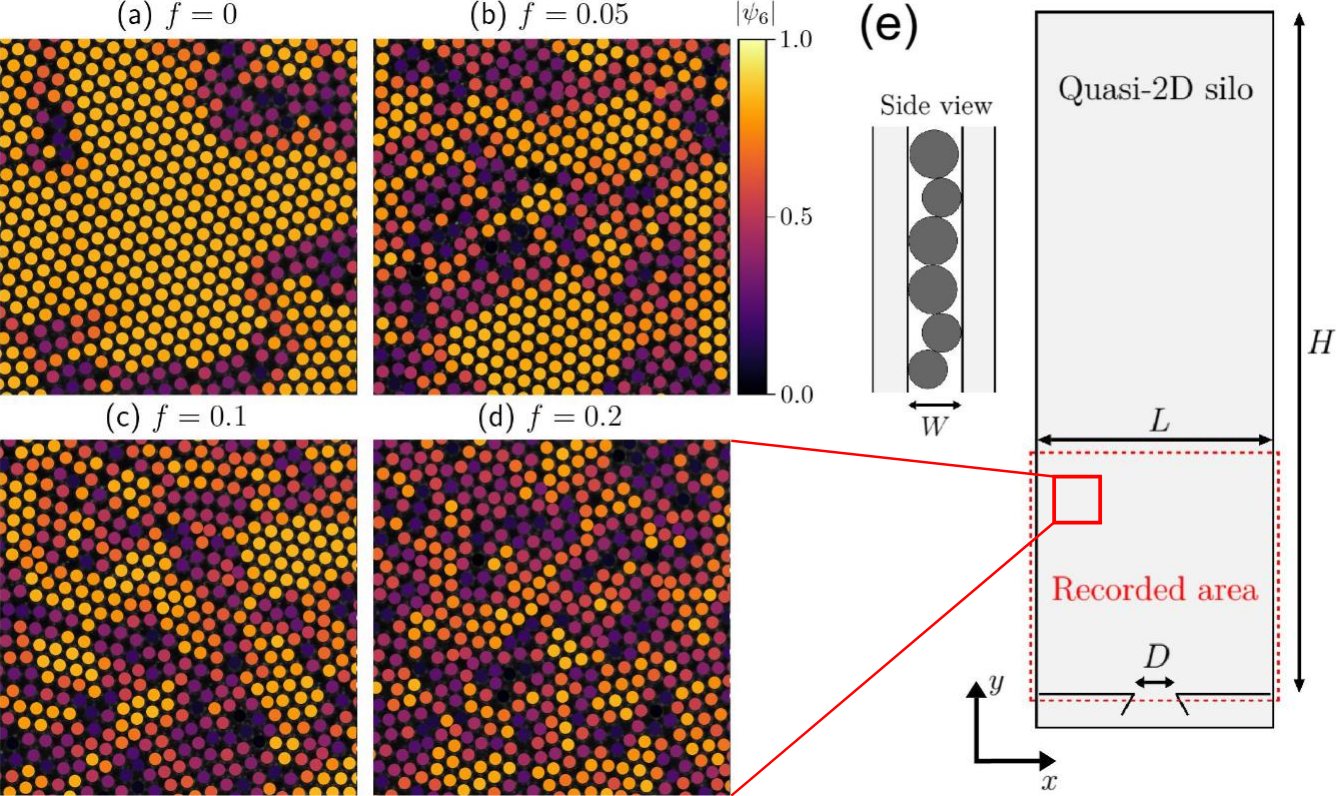}
    \caption{Four snapshots of the bidisperse granular media in a discharging 2D silo for different values of the fraction (a) $f=0$, (b) $f=0.05$, (c) $f=0.1$ and (d) $f=0.2$. Each particle is colored according to the hexatic order parameter modulus $|\psi_6|$. (e) Sketch of the experimental quasi-2D silo with relevant parameters.}
    \label{Fig:crystallization}
\end{figure}

Granular materials, composed of discrete macroscopic particles, display a wide range of collective behaviors that often defy intuition, owing to their athermal nature and dissipative interactions. Among these behaviors, structural ordering is a central aspect. When monodisperse spherical grains are densely packed, they tend to self-organize into highly ordered configurations, such as hcp and fcc lattices, which represent efficient packing arrangements \cite{torquato2010jammed}. However, even a slight degree of size polydispersity can disrupt this tendency, preventing crystallization and promoting disordered or amorphous structures that lack long-range translational order \cite{watanabe2008direct, kondic2012topology}. This effect is particularly pronounced in two-dimensional systems, where monodisperse disks spontaneously form hexagonal crystallites under confinement (see Fig.~\ref{Fig:crystallization}(a)).

While it is well established that stresses in granular media propagate along discrete contact networks, the influence of structural order on this transmission, particularly under flow conditions, remains a complex and actively debated topic \cite{reis2006crystallization, watanabe2008direct, benyamine2014discharge, downs2021topographic, bai2023mesoscopic, carlevaro2024flow}. To suppress undesired ordering effects, especially in experimental investigations of flow, compaction, or jamming, bidisperse mixtures are commonly used. This strategy prompts several fundamental questions: which combinations of grain size ratios and mixing fractions most effectively hinder crystallization, and to what extent does the presence or absence of local structural order modify the mechanical response and dynamical behavior of granular systems?

The quasi-two-dimensional geometry of the silo provides an ideal framework to investigate the interplay between particle ordering and granular flow. Over the past decades, granular discharge from silos has been extensively studied, leading to the development of several semi-empirical models that capture key flow properties such as the discharge rate \cite{Hagen1852, beverloo1961flow, janda2012flow}, the internal velocity field \cite{litwiniszyn1964application, mullins1974experimental, nedderman1979kinematic, medina1998velocity, choi2005velocity}, and the dynamics near the outlet \cite{janda2012flow, benyamine2017discharge}. Despite this knowledge, the influence of structural ordering, particularly crystallization, within the bulk reservoir of the silo remains largely unexplored. This gap motivates a detailed investigation of how local particle arrangements affect the flow behavior in bidisperse systems.

We designed a quasi-two-dimensional flat-bottomed silo that constrains the grains to a single monolayer (see Fig.~\ref{Fig:crystallization}(e)). The front and back walls are made of polycarbonate coated with a conductive layer to mitigate triboelectric charging. The cell dimensions are $H = 300$~mm (height), $L = 100$~mm (width), and $W = 1.25$~mm (depth), with a fixed outlet width of $18$~mm. The granular medium consists of a binary mixture of steel spheres with diameters $d_1 = 1.0$~mm and $d_2 = 1.2$~mm, and we define the average reference diameter as $d = 1.1$~mm. The mass fraction of small beads is denoted by $f$, representing the proportion of $1.0$~mm particles in the mixture. Once the aperture is opened, by lifting a retaining wire, grains begin to flow out, allowing us to study the discharge dynamics as a function of $f$. For each selected mass fraction, five independent discharge experiments were performed. The flow was recorded using a high-speed camera operating at 2000 frames per second. Each video sequence, captured with a resolution of $1024 \times 1024$ pixels (corresponding to a physical area of approximately $91d \times 91d$), focuses on the bottom region of the silo during the steady-state phase of the discharge. The Lagrangian quantities associated with each grain are projected onto an Eulerian grid composed of square cells of size $d \times d$. The grid is constructed such that its bottom-central cell aligns with the center of the silo outlet. Within this framework, Eulerian fields for any measurable quantity can be obtained by averaging the corresponding Lagrangian values of all particles intersecting a given cell at position $(x, y)$.

Figure \ref{Fig:crystallization}(a–d) presents snapshots of the granular packings within the silo for four different values of $f$, ranging from the monodisperse case ($f=0$) to a low fraction of small particles ($f=0.2$). Grains are color-coded according to their local structural environment, from crystalline regions (yellow) to amorphous ones (dark). To quantify the local order, we compute for each particle $k$ the hexatic order parameter based on the angular positions $\theta_{k\ell}$ of its $N_k$ nearest neighbors
\begin{equation}
\psi_{6k} = \frac{1}{N_k} \sum_{\ell=1}^{N_k} \exp\left(i6\theta_{k\ell}\right),
\label{Eq:hexatic_parameter}
\end{equation}
where $|\psi_{6k}|$ close to $1$ indicates a strong sixfold symmetry typical of hexagonal packing. The images clearly show that the large crystalline domains observed at $f=0$ are rapidly disrupted as $f$ increases, giving way to smaller clusters and disordered regions. This confirms that bidispersity suppresses long-range order and promotes amorphous configurations. 

\begin{figure}[h!]
    \centering
    \includegraphics[width=0.43\textwidth]{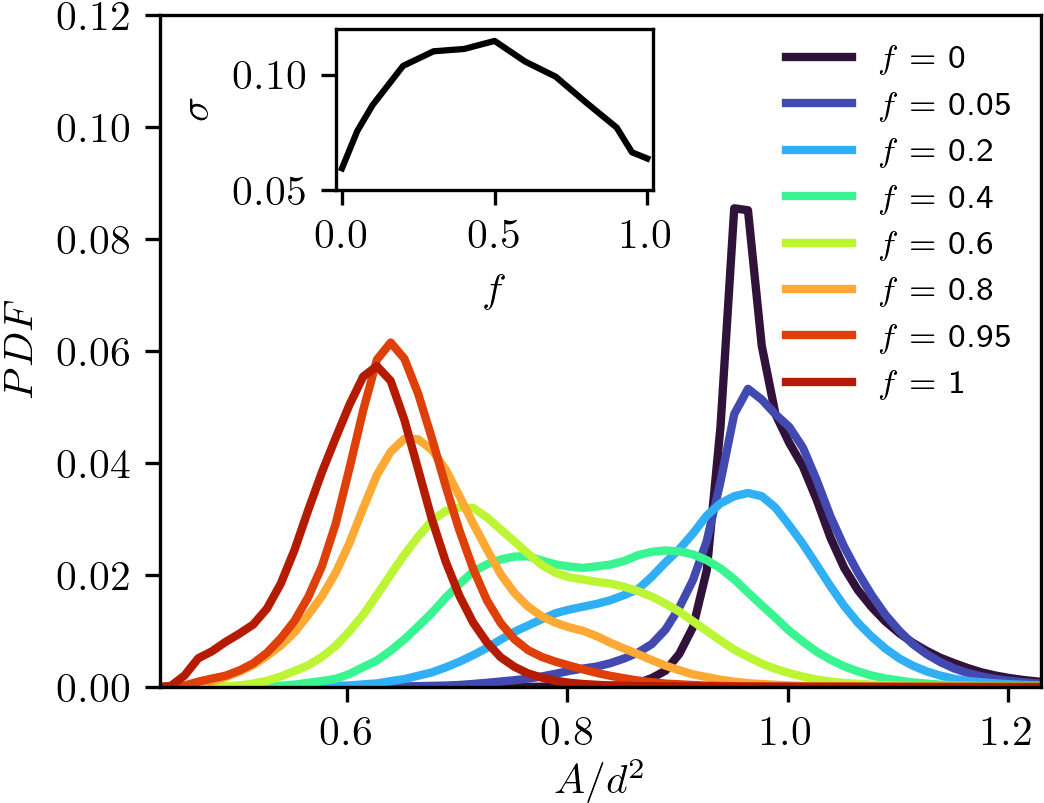}
    \vskip -0.3cm
    \caption{Probability Distribution Function (PDF) of Voronoi cell areas $A$ normalized by a typical grain area $d^2$. Different mixing fractions $f$ are shown exhibiting different patterns. The inset shows the standard deviation $\sigma$ of each distribution as a function of $f$.}
    \label{Fig:pdf_voro}
\end{figure}

A second structural measure confirms these observations. A Voronoi tessellation is applied to all grains, assigning to each grain $k$ a Voronoi cell of area $A_k$. We compute the Probability Distribution Function (PDF) of the dimensionless areas $A/d^2$ for various values of $f$, as shown in Figure~\ref{Fig:pdf_voro}. For the monodisperse case ($f=0$), only large grains are present and the distribution is sharply peaked. As $f$ increases, a second peak emerges at smaller normalized areas, indicating the inclusion of smaller grains. The overall distribution broadens, reaching its widest spread near $f = 0.5$, before narrowing again for high $f$ values. At $f=1$, where only small particles are present, crystalline domains reappear, although the peak remains broader than in the $f=0$ case. This is due to the limited confinement in the silo's geometry, allowing some particle overlap and thus reducing positional order. The inset of Figure~\ref{Fig:pdf_voro} shows the standard deviation $\sigma$ of the PDFs, which reaches a maximum around $f=0.5$, consistent with the regime of minimal crystalline order.

\begin{figure}[h!]
    \centering
    \includegraphics[width=0.43\textwidth]{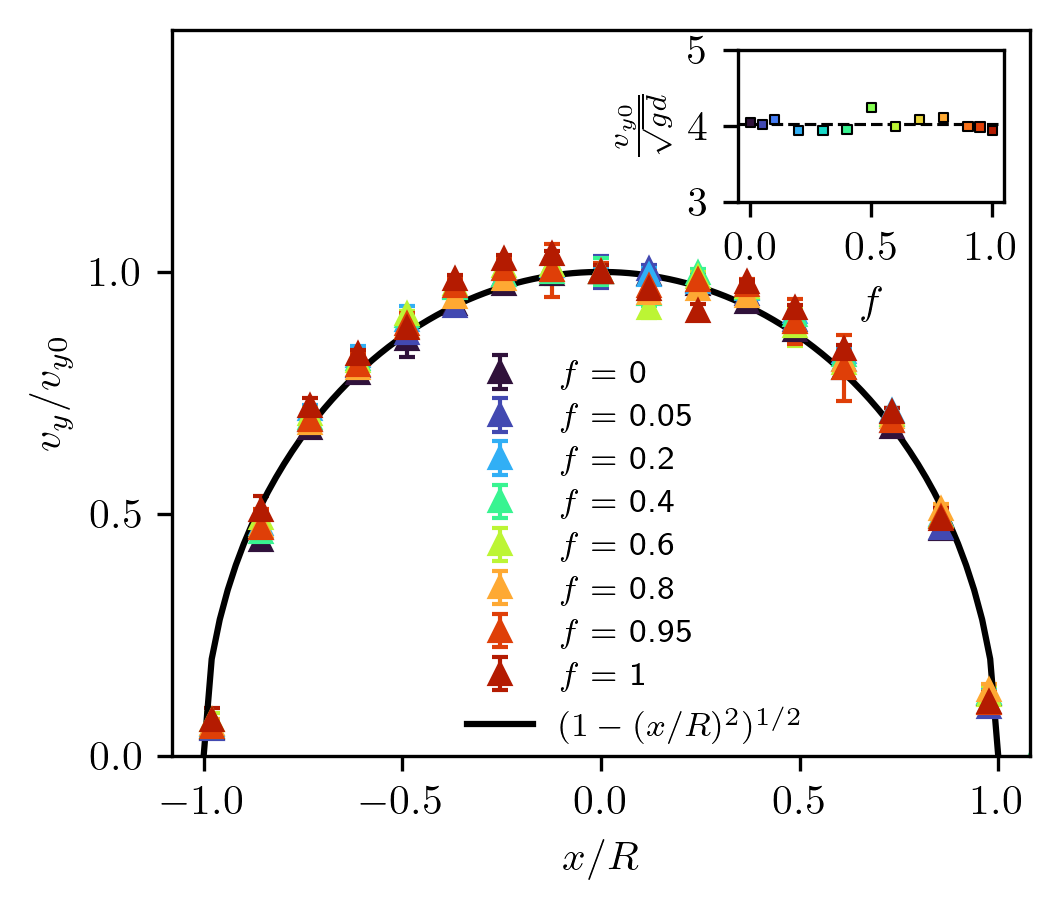}
    \vskip -0.5cm
    \caption{Normalized vertical velocity $v_y/v_{y0}$ profiles for all mixing fractions $f$. The black line corresponds to the fitting $\left( 1 - (x/R)^2\right)^{1/2}$ with $R$ the radius of the silo aperture. The inset presents the central velocity $v_{y0}$ normalized by the characteristic speed  $\sqrt{gd}$, emphasizing no significant dependency of the mixing fractions $f$.}
    \label{Fig:velocity_profiles}
\end{figure}

The above results are confirming the validity of most experimental protocols in which binary mixtures are used to avoid crystallization. The next relevant question is to measure the effect of this mixture on the global granular flow. In particular, we compute the average vertical velocity field $v_y(x, y)$ as a function of the mixing fraction $f$. Let us start by the outlet of the silo where the velocity profiles exhibit no significant dependence on $f$, in agreement with prior observations reported in \cite{janda2012flow, carlevaro2024flow} (see Fig. \ref{Fig:velocity_profiles}). The expected profile around the outlet is fitted over all data. This indicates that orientational correlations have little influence in this dilute region of the flow, where the granular medium is subject to a strong pressure drop \cite{perge2012evolution}. 

\begin{figure*}[ht]
    \centering
    \includegraphics[width=1\textwidth]{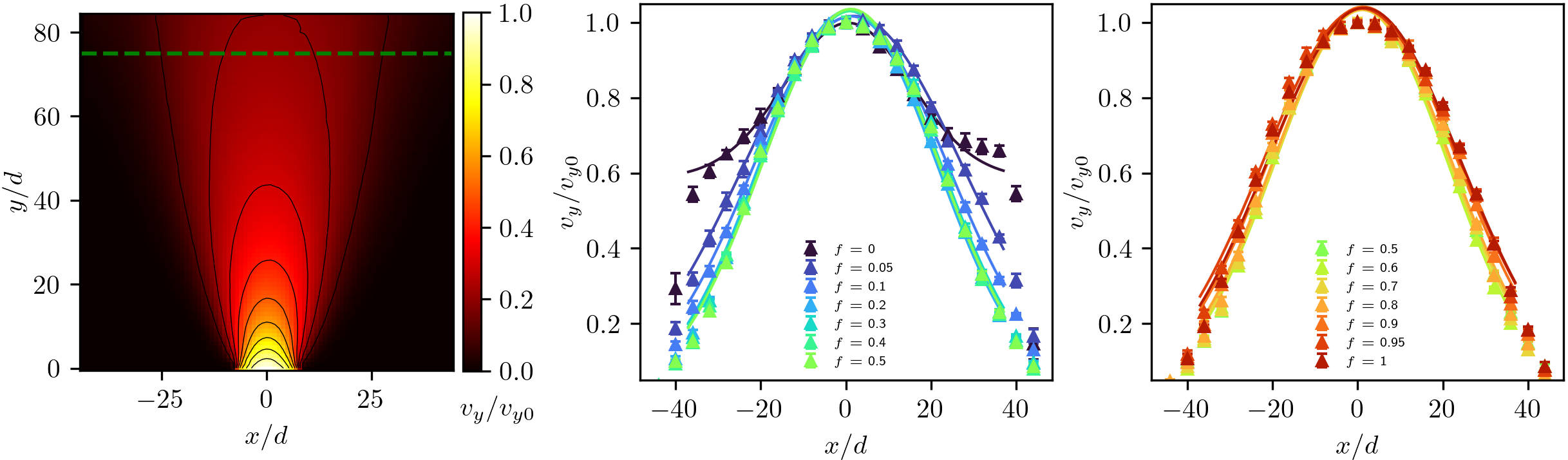}
    \caption{(a) Time-average field of the normalized vertical velocity field $v_y / v_{y0}$ for $f=0.5$. Bulk velocity profiles for (b) $f \in [0, 0.5]$ and (c) $f \in [0.5, 1]$ along the horizontal dashed line at $y/d=75$ represented in green on (a). All solid curves corresponds to the fit of Eq.(\ref{Eq:kinematic_model}) considering a plug flow component.} 
    \label{Fig:gaussian_profiles}
\end{figure*}

However, the effect of mixing granular species can be seen in the silo far from the outlet. A typical velocity field $v(x,y)$ is shown in Fig \ref{Fig:gaussian_profiles}(a) where dead zones are in dark while fast moving parts are in yellow close to the outlet. A large part of the silo is moving and this is well described by the kinematic model introduced by Nedderman and Tüzün~\cite{nedderman1979kinematic}, based on the earlier work of Litwiniszyn~\cite{litwiniszyn1964application}. The model assumes that the horizontal velocity is proportional to the horizontal gradient of the vertical velocity, i.e. $v_x=\partial_x v_y$. This leads to Gaussian-like-shaped vertical velocity profiles of the form 
\begin{equation}
v_y(x,y)\,=\,\frac{Q}{\sqrt{4 \pi b y}} \, \exp \left( \frac{-x^2}{4 b y} \right),
\label{Eq:kinematic_model}
\end{equation}
where $Q$ is the time-averaged flow rate. The model introduces a single fitting parameter, $b$, which represents a characteristic diffusion length of the medium \cite{choi2005velocity}. These velocity profiles are shown in Fig \ref{Fig:gaussian_profiles}(b,c) for respectively low $f$ and high $f$ values. The profile is taken at the same height $y/d=75$ for comparison. For $f=0$, the velocity profile exhibits a uniform component, which corresponds to a crystalline block whose dislocation threshold is significantly higher than the shear stress present in this region. One concludes that velocity field $v(x,y)$ is altered by the onset of the crystallization phenomenon. 

\begin{figure}[h!]
    \centering
    \includegraphics[width=0.48\textwidth]{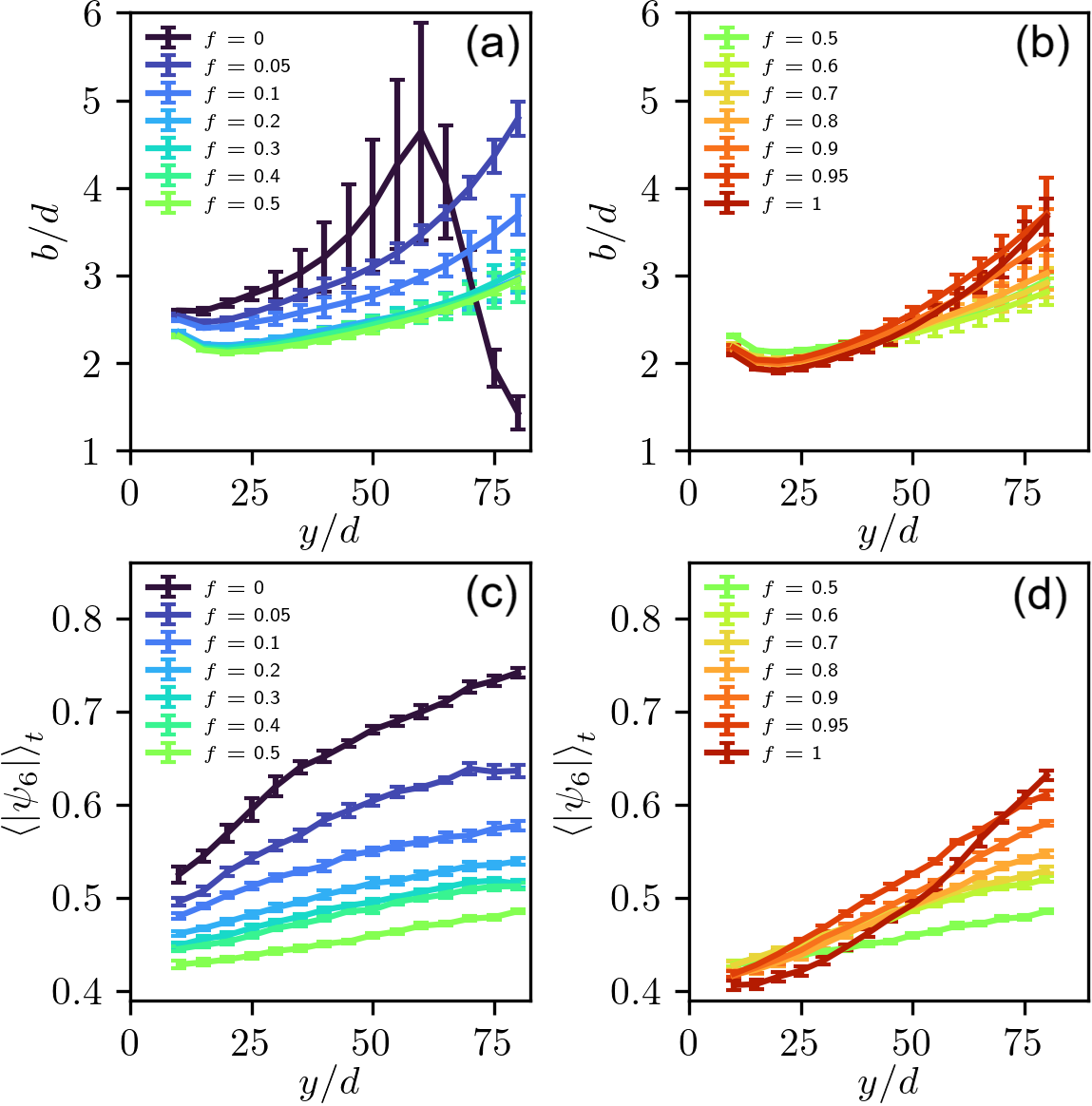}
    \caption{Evolution of the normalized parameter $b/d$ in the silo for (a) $f \in [0, 0.5]$ and (b) $f \in [0.5, 1]$. Modulus of the local hexatic parameter order $\left<|\psi_6|\right>_t$ as a function of height for (c) $f \in [0, 0.5]$ and (d) $f \in [0.5, 1]$.}
    \label{Fig:kinematic_model}
\end{figure}

Removing the plug flow component, the Gaussian-like shape can still be fitted using the kinematic model, which allows the characteristic diffusion length of the medium $b$ to be extracted as a function of the height $y/d$ and the mixing fraction $f$. Figures \ref{Fig:kinematic_model} (a) and (b) show resulting $b/d$ values as a function of the vertical position $y/d$. Distinguishing $f<0.5$ and $f>0.5$ allows to evidence that the diffusion length $b$ is minimal for $f\approx 0.5$ while it increases significantly when ordering is present. The diffusion length $b$ systematically increases with height within the silo. This trend is consistent with the observations of Medina et al., who performed measurements in a monodisperse granular medium, under conditions where the aspect ratio between the silo gap and particle size corresponds to our case with $f=1$ \cite{medina1998velocity}.

In this study, we further demonstrate that the increase in diffusion length becomes even more pronounced when crystallization occurs within the granular medium. Specifically, the diffusion length reaches a value of $b/d=4.5$ in the most crystalline case ($f<0.05$), which exceeds previously reported values for slightly polydisperse granular systems \cite{mullins1974experimental, samadani1999size, medina1998velocity, chen2021measurement}.

\newpage

This enhanced diffusivity is strongly correlated with the cohesion induced by crystallization. The formation of crystalline blocks at the mesoscopic scale increases the number of mechanically stable configurations available to the grains, thereby reducing the prevalence of microscopically frustrated packings. As a consequence, the likelihood of localized plastic events occurring during flow decreases—a mechanism suggested to underlie the anomalous diffusion behavior observed in granular materials \cite{zuriguel2019velocity}.

To explore this hypothesis, we compute the local time-averaged magnitude of the hexatic order parameter, $\left<| \psi_6|\right>_t$, as an Eulerian field. Previous studies have established a link between dynamic heterogeneity and short-range orientational order \cite{watanabe2008direct}. While instantaneous $|\psi_6|$ values reflect the presence of hexatic structures at a given time, the time-averaged field $\left< |\psi_6|\right>_t$ is inversely proportional to the frequency of plastic rearrangements at each location.

Our results show that $\left< |\psi_6|\right>_t$ varies only slightly along the horizontal axis $x$, but increases significantly with height, as depicted in Figure \ref{Fig:kinematic_model}. In regions of high velocity, $\left< |\psi_6|\right>_t$ ranges between $0.4$ and $0.55$, indicating substantial dislocation activity consistent with frequent plastic events. Conversely, higher in the silo, this parameter can reach values up to $0.75$ in the fully crystalline case ($f=0$), suggesting slower dynamics in these low-velocity regions where plastic events are suppressed by both elevated pressure \cite{walker1966approximate} and the crystallinity-induced cohesion identified here. Velocity profiles at the outlet and in the bulk are largely similar, except when $\left<| \psi_6|\right>_t$ exceeds the threshold of $0.6$. This critical threshold was previously identified in experimental studies of two-dimensional vibrated granular media \cite{downs2021topographic}. Above this value, the velocity profiles develop broader tails, which are well captured by the diffusion length parameter $b$.

\begin{figure}[h!]
    \centering
    \includegraphics[width=0.37\textwidth]{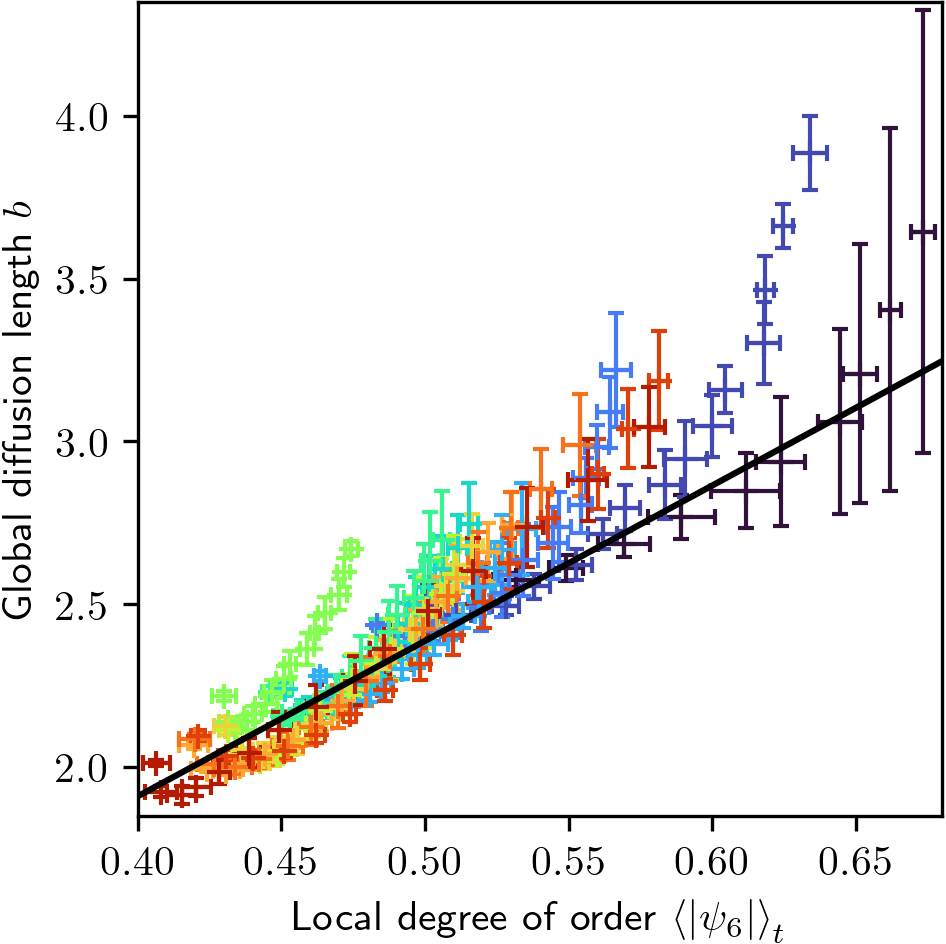}
    \caption{Correlation between the local modulus of the hexatic parameter order $\left<|\psi_6|\right>_t$ and the global caracteristic diffusion length $b$ ($r=0.942$) in the range $y/d \in[10, 70]$. The black line corresponds to a line with slope $4.8$.}
    \label{Fig:correlation}
\end{figure}

A clear correlation thus emerges between the global diffusion length $b$ and the local time-averaged degree of order $\left<| \psi_6|\right>_t$, as illustrated in Figure \ref{Fig:correlation}. A high rate of dislocations, reflected by low $\left< |\psi_6|\right>_t$ values, results in poor diffusion properties of the granular assembly, as momentum transfer is hindered by energy dissipation from plastic events. In contrast, a low dislocation rate—indicated by high $\left< |\psi_6|\right>_t$ values—is associated with more efficient momentum propagation. Furthermore, this mechanism is reinforced by increasing confining pressure, potentially explaining the rise in the global diffusion parameter $b$ observed in previous studies \cite{medina1998velocity}.

In summary, we investigated how local structural ordering, i.e. crystallization, affects the flow dynamics of bidisperse granular materials in a quasi-2D silo. We demonstrated that the presence of crystalline domains enhances the local rigidity of the medium, thereby facilitating momentum transfer and leading to a measurable increase in the diffusion length $b$ within the kinematic model. A strong correlation was established between the hexatic order parameter and $b$, highlighting the pivotal role of microstructural organization in governing macroscopic flow behavior. Furthermore, we showed that even in the absence of significant crystallization, the pressure increase with height promotes the stabilization of local orientational order, which in turn causes a gradual increase in $b$. These findings point to pressure as a key factor mediating the interplay between structural order and transport properties in granular silo flows.\\

This work was carried out as part of [optionally specify the project or call], with the financial and scientific support of PIA project Lorraine Université d'Excellence (LUE), which significantly contributed to the achievement of this research (2025).

\bibliography{_biblio}

\begin{thebibliography}{22}%
\makeatletter
\providecommand \@ifxundefined [1]{%
 \@ifx{#1\undefined}
}%
\providecommand \@ifnum [1]{%
 \ifnum #1\expandafter \@firstoftwo
 \else \expandafter \@secondoftwo
 \fi
}%
\providecommand \@ifx [1]{%
 \ifx #1\expandafter \@firstoftwo
 \else \expandafter \@secondoftwo
 \fi
}%
\providecommand \natexlab [1]{#1}%
\providecommand \enquote  [1]{``#1''}%
\providecommand \bibnamefont  [1]{#1}%
\providecommand \bibfnamefont [1]{#1}%
\providecommand \citenamefont [1]{#1}%
\providecommand \href@noop [0]{\@secondoftwo}%
\providecommand \href [0]{\begingroup \@sanitize@url \@href}%
\providecommand \@href[1]{\@@startlink{#1}\@@href}%
\providecommand \@@href[1]{\endgroup#1\@@endlink}%
\providecommand \@sanitize@url [0]{\catcode `\\12\catcode `\$12\catcode
  `\&12\catcode `\#12\catcode `\^12\catcode `\_12\catcode `\%12\relax}%
\providecommand \@@startlink[1]{}%
\providecommand \@@endlink[0]{}%
\providecommand \url  [0]{\begingroup\@sanitize@url \@url }%
\providecommand \@url [1]{\endgroup\@href {#1}{\urlprefix }}%
\providecommand \urlprefix  [0]{URL }%
\providecommand \Eprint [0]{\href }%
\providecommand \doibase [0]{https://doi.org/}%
\providecommand \selectlanguage [0]{\@gobble}%
\providecommand \bibinfo  [0]{\@secondoftwo}%
\providecommand \bibfield  [0]{\@secondoftwo}%
\providecommand \translation [1]{[#1]}%
\providecommand \BibitemOpen [0]{}%
\providecommand \bibitemStop [0]{}%
\providecommand \bibitemNoStop [0]{.\EOS\space}%
\providecommand \EOS [0]{\spacefactor3000\relax}%
\providecommand \BibitemShut  [1]{\csname bibitem#1\endcsname}%
\let\auto@bib@innerbib\@empty
\bibitem [{\citenamefont {Torquato}\ and\ \citenamefont
  {Stillinger}(2010)}]{torquato2010jammed}%
  \BibitemOpen
  \bibfield  {author} {\bibinfo {author} {\bibfnamefont {S.}~\bibnamefont
  {Torquato}}\ and\ \bibinfo {author} {\bibfnamefont {F.~H.}\ \bibnamefont
  {Stillinger}},\ }\bibfield  {title} {\bibinfo {title} {Jammed hard-particle
  packings: From kepler to bernal and beyond},\ }\href@noop {} {\bibfield
  {journal} {\bibinfo  {journal} {Reviews of modern physics}\ }\textbf
  {\bibinfo {volume} {82}},\ \bibinfo {pages} {2633} (\bibinfo {year}
  {2010})}\BibitemShut {NoStop}%
\bibitem [{\citenamefont {Watanabe}\ and\ \citenamefont
  {Tanaka}(2008)}]{watanabe2008direct}%
  \BibitemOpen
  \bibfield  {author} {\bibinfo {author} {\bibfnamefont {K.}~\bibnamefont
  {Watanabe}}\ and\ \bibinfo {author} {\bibfnamefont {H.}~\bibnamefont
  {Tanaka}},\ }\bibfield  {title} {\bibinfo {title} {Direct observation of
  medium-range crystalline order in granular liquids near the glass
  transition},\ }\href@noop {} {\bibfield  {journal} {\bibinfo  {journal}
  {Physical review letters}\ }\textbf {\bibinfo {volume} {100}},\ \bibinfo
  {pages} {158002} (\bibinfo {year} {2008})}\BibitemShut {NoStop}%
\bibitem [{\citenamefont {Kondic}\ \emph {et~al.}(2012)\citenamefont {Kondic},
  \citenamefont {Goullet}, \citenamefont {O'Hern}, \citenamefont {Kramar},
  \citenamefont {Mischaikow},\ and\ \citenamefont
  {Behringer}}]{kondic2012topology}%
  \BibitemOpen
  \bibfield  {author} {\bibinfo {author} {\bibfnamefont {L.}~\bibnamefont
  {Kondic}}, \bibinfo {author} {\bibfnamefont {A.}~\bibnamefont {Goullet}},
  \bibinfo {author} {\bibfnamefont {C.}~\bibnamefont {O'Hern}}, \bibinfo
  {author} {\bibfnamefont {M.}~\bibnamefont {Kramar}}, \bibinfo {author}
  {\bibfnamefont {K.}~\bibnamefont {Mischaikow}},\ and\ \bibinfo {author}
  {\bibfnamefont {R.}~\bibnamefont {Behringer}},\ }\bibfield  {title} {\bibinfo
  {title} {Topology of force networks in compressed granular media},\
  }\href@noop {} {\bibfield  {journal} {\bibinfo  {journal} {Europhysics
  Letters}\ }\textbf {\bibinfo {volume} {97}},\ \bibinfo {pages} {54001}
  (\bibinfo {year} {2012})}\BibitemShut {NoStop}%
\bibitem [{\citenamefont {Reis}\ \emph {et~al.}(2006)\citenamefont {Reis},
  \citenamefont {Ingale},\ and\ \citenamefont
  {Shattuck}}]{reis2006crystallization}%
  \BibitemOpen
  \bibfield  {author} {\bibinfo {author} {\bibfnamefont {P.~M.}\ \bibnamefont
  {Reis}}, \bibinfo {author} {\bibfnamefont {R.~A.}\ \bibnamefont {Ingale}},\
  and\ \bibinfo {author} {\bibfnamefont {M.~D.}\ \bibnamefont {Shattuck}},\
  }\bibfield  {title} {\bibinfo {title} {Crystallization of a
  quasi-two-dimensional granular fluid},\ }\href@noop {} {\bibfield  {journal}
  {\bibinfo  {journal} {Physical review letters}\ }\textbf {\bibinfo {volume}
  {96}},\ \bibinfo {pages} {258001} (\bibinfo {year} {2006})}\BibitemShut
  {NoStop}%
\bibitem [{\citenamefont {Benyamine}\ \emph {et~al.}(2014)\citenamefont
  {Benyamine}, \citenamefont {Djermane}, \citenamefont {Dalloz-Dubrujeaud},\
  and\ \citenamefont {Aussillous}}]{benyamine2014discharge}%
  \BibitemOpen
  \bibfield  {author} {\bibinfo {author} {\bibfnamefont {M.}~\bibnamefont
  {Benyamine}}, \bibinfo {author} {\bibfnamefont {M.}~\bibnamefont {Djermane}},
  \bibinfo {author} {\bibfnamefont {B.}~\bibnamefont {Dalloz-Dubrujeaud}},\
  and\ \bibinfo {author} {\bibfnamefont {P.}~\bibnamefont {Aussillous}},\
  }\bibfield  {title} {\bibinfo {title} {Discharge flow of a bidisperse
  granular media from a silo},\ }\href@noop {} {\bibfield  {journal} {\bibinfo
  {journal} {Physical Review E}\ }\textbf {\bibinfo {volume} {90}},\ \bibinfo
  {pages} {032201} (\bibinfo {year} {2014})}\BibitemShut {NoStop}%
\bibitem [{\citenamefont {Downs}\ \emph {et~al.}(2021)\citenamefont {Downs},
  \citenamefont {Smith}, \citenamefont {Mandadapu}, \citenamefont {Garrahan},\
  and\ \citenamefont {Smith}}]{downs2021topographic}%
  \BibitemOpen
  \bibfield  {author} {\bibinfo {author} {\bibfnamefont {J.}~\bibnamefont
  {Downs}}, \bibinfo {author} {\bibfnamefont {N.}~\bibnamefont {Smith}},
  \bibinfo {author} {\bibfnamefont {K.}~\bibnamefont {Mandadapu}}, \bibinfo
  {author} {\bibfnamefont {J.}~\bibnamefont {Garrahan}},\ and\ \bibinfo
  {author} {\bibfnamefont {M.}~\bibnamefont {Smith}},\ }\bibfield  {title}
  {\bibinfo {title} {Topographic control of order in quasi-2d granular phase
  transitions},\ }\href@noop {} {\bibfield  {journal} {\bibinfo  {journal}
  {Physical Review Letters}\ }\textbf {\bibinfo {volume} {127}},\ \bibinfo
  {pages} {268002} (\bibinfo {year} {2021})}\BibitemShut {NoStop}%
\bibitem [{\citenamefont {Bai}\ \emph {et~al.}(2023)\citenamefont {Bai},
  \citenamefont {Li}, \citenamefont {Hong}, \citenamefont {Pan},\ and\
  \citenamefont {Fei}}]{bai2023mesoscopic}%
  \BibitemOpen
  \bibfield  {author} {\bibinfo {author} {\bibfnamefont {J.}~\bibnamefont
  {Bai}}, \bibinfo {author} {\bibfnamefont {J.}~\bibnamefont {Li}}, \bibinfo
  {author} {\bibfnamefont {G.}~\bibnamefont {Hong}}, \bibinfo {author}
  {\bibfnamefont {J.}~\bibnamefont {Pan}},\ and\ \bibinfo {author}
  {\bibfnamefont {H.}~\bibnamefont {Fei}},\ }\bibfield  {title} {\bibinfo
  {title} {Mesoscopic evolution and kinetic properties of dense granular flow
  crystallization under continuous shear induction},\ }\href@noop {} {\bibfield
   {journal} {\bibinfo  {journal} {Powder Technology}\ }\textbf {\bibinfo
  {volume} {426}},\ \bibinfo {pages} {118615} (\bibinfo {year}
  {2023})}\BibitemShut {NoStop}%
\bibitem [{\citenamefont {Carlevaro}\ \emph {et~al.}(2024)\citenamefont
  {Carlevaro}, \citenamefont {Kozlowski},\ and\ \citenamefont
  {Pugnaloni}}]{carlevaro2024flow}%
  \BibitemOpen
  \bibfield  {author} {\bibinfo {author} {\bibfnamefont {C.~M.}\ \bibnamefont
  {Carlevaro}}, \bibinfo {author} {\bibfnamefont {R.}~\bibnamefont
  {Kozlowski}},\ and\ \bibinfo {author} {\bibfnamefont {L.~A.}\ \bibnamefont
  {Pugnaloni}},\ }\bibfield  {title} {\bibinfo {title} {Flow rate in 2d silo
  discharge of binary granular mixtures: the role of ordering in monosized
  systems},\ }\href@noop {} {\bibfield  {journal} {\bibinfo  {journal}
  {Frontiers in Soft Matter}\ }\textbf {\bibinfo {volume} {4}},\ \bibinfo
  {pages} {1340744} (\bibinfo {year} {2024})}\BibitemShut {NoStop}%
\bibitem [{\citenamefont {Hagen}(1852)}]{Hagen1852}%
  \BibitemOpen
  \bibfield  {author} {\bibinfo {author} {\bibfnamefont {G.}~\bibnamefont
  {Hagen}},\ }\bibfield  {title} {\bibinfo {title} {{\"U}ber den {D}ruck und
  die {B}ewegung des trocknen {S}andes},\ }\href@noop {} {\bibfield  {journal}
  {\bibinfo  {journal} {Bericht {\"u}ber die zur {B}ekanntmachung geeigneten
  {V}erhandlungen der {K}{\"o}niglich Preussischen Akademie der Wissenschaften
  zu Berlin}\ ,\ \bibinfo {pages} {35}} (\bibinfo {year} {1852})}\BibitemShut
  {NoStop}%
\bibitem [{\citenamefont {Beverloo}\ \emph {et~al.}(1961)\citenamefont
  {Beverloo}, \citenamefont {Leniger},\ and\ \citenamefont {Van~de
  Velde}}]{beverloo1961flow}%
  \BibitemOpen
  \bibfield  {author} {\bibinfo {author} {\bibfnamefont {W.~A.}\ \bibnamefont
  {Beverloo}}, \bibinfo {author} {\bibfnamefont {H.~A.}\ \bibnamefont
  {Leniger}},\ and\ \bibinfo {author} {\bibfnamefont {J.}~\bibnamefont {Van~de
  Velde}},\ }\bibfield  {title} {\bibinfo {title} {The flow of granular solids
  through orifices},\ }\href@noop {} {\bibfield  {journal} {\bibinfo  {journal}
  {Chemical engineering science}\ }\textbf {\bibinfo {volume} {15}},\ \bibinfo
  {pages} {260} (\bibinfo {year} {1961})}\BibitemShut {NoStop}%
\bibitem [{\citenamefont {Janda}\ \emph {et~al.}(2012)\citenamefont {Janda},
  \citenamefont {Zuriguel},\ and\ \citenamefont {Maza}}]{janda2012flow}%
  \BibitemOpen
  \bibfield  {author} {\bibinfo {author} {\bibfnamefont {A.}~\bibnamefont
  {Janda}}, \bibinfo {author} {\bibfnamefont {I.}~\bibnamefont {Zuriguel}},\
  and\ \bibinfo {author} {\bibfnamefont {D.}~\bibnamefont {Maza}},\ }\bibfield
  {title} {\bibinfo {title} {Flow rate of particles through apertures obtained
  from self-similar density and velocity profiles},\ }\href@noop {} {\bibfield
  {journal} {\bibinfo  {journal} {Physical review letters}\ }\textbf {\bibinfo
  {volume} {108}},\ \bibinfo {pages} {248001} (\bibinfo {year}
  {2012})}\BibitemShut {NoStop}%
\bibitem [{\citenamefont {Litwiniszyn}(1964)}]{litwiniszyn1964application}%
  \BibitemOpen
  \bibfield  {author} {\bibinfo {author} {\bibfnamefont {J.}~\bibnamefont
  {Litwiniszyn}},\ }\bibfield  {title} {\bibinfo {title} {An application of the
  random walk argument to the mechanics of granular media},\ }\href@noop {}
  {\bibfield  {journal} {\bibinfo  {journal} {Rheology and Soil Mechanics:
  Symposium Grenoble, April 1--8, 1964}\ ,\ \bibinfo {pages} {82}} (\bibinfo
  {year} {1964})}\BibitemShut {NoStop}%
\bibitem [{\citenamefont {Mullins}(1974)}]{mullins1974experimental}%
  \BibitemOpen
  \bibfield  {author} {\bibinfo {author} {\bibfnamefont {W.}~\bibnamefont
  {Mullins}},\ }\bibfield  {title} {\bibinfo {title} {Experimental evidence for
  the stochastic theory of particle flow under gravity},\ }\href@noop {}
  {\bibfield  {journal} {\bibinfo  {journal} {Powder Technology}\ }\textbf
  {\bibinfo {volume} {9}},\ \bibinfo {pages} {29} (\bibinfo {year}
  {1974})}\BibitemShut {NoStop}%
\bibitem [{\citenamefont {Nedderman}\ and\ \citenamefont
  {T{\"u}z{\"u}n}(1979)}]{nedderman1979kinematic}%
  \BibitemOpen
  \bibfield  {author} {\bibinfo {author} {\bibfnamefont {R.}~\bibnamefont
  {Nedderman}}\ and\ \bibinfo {author} {\bibfnamefont {U.}~\bibnamefont
  {T{\"u}z{\"u}n}},\ }\bibfield  {title} {\bibinfo {title} {A kinematic model
  for the flow of granular materials},\ }\href@noop {} {\bibfield  {journal}
  {\bibinfo  {journal} {Powder Technology}\ }\textbf {\bibinfo {volume} {22}},\
  \bibinfo {pages} {243} (\bibinfo {year} {1979})}\BibitemShut {NoStop}%
\bibitem [{\citenamefont {Medina}\ \emph {et~al.}(1998)\citenamefont {Medina},
  \citenamefont {Cordova}, \citenamefont {Luna},\ and\ \citenamefont
  {Trevino}}]{medina1998velocity}%
  \BibitemOpen
  \bibfield  {author} {\bibinfo {author} {\bibfnamefont {A.}~\bibnamefont
  {Medina}}, \bibinfo {author} {\bibfnamefont {J.}~\bibnamefont {Cordova}},
  \bibinfo {author} {\bibfnamefont {E.}~\bibnamefont {Luna}},\ and\ \bibinfo
  {author} {\bibfnamefont {C.}~\bibnamefont {Trevino}},\ }\bibfield  {title}
  {\bibinfo {title} {Velocity field measurements in granular gravity flow in a
  near 2d silo},\ }\href@noop {} {\bibfield  {journal} {\bibinfo  {journal}
  {Physics Letters A}\ }\textbf {\bibinfo {volume} {250}},\ \bibinfo {pages}
  {111} (\bibinfo {year} {1998})}\BibitemShut {NoStop}%
\bibitem [{\citenamefont {Choi}\ \emph {et~al.}(2005)\citenamefont {Choi},
  \citenamefont {Kudrolli},\ and\ \citenamefont {Bazant}}]{choi2005velocity}%
  \BibitemOpen
  \bibfield  {author} {\bibinfo {author} {\bibfnamefont {J.}~\bibnamefont
  {Choi}}, \bibinfo {author} {\bibfnamefont {A.}~\bibnamefont {Kudrolli}},\
  and\ \bibinfo {author} {\bibfnamefont {M.~Z.}\ \bibnamefont {Bazant}},\
  }\bibfield  {title} {\bibinfo {title} {Velocity profile of granular flows
  inside silos and hoppers},\ }\href@noop {} {\bibfield  {journal} {\bibinfo
  {journal} {Journal of Physics: Condensed Matter}\ }\textbf {\bibinfo {volume}
  {17}},\ \bibinfo {pages} {S2533} (\bibinfo {year} {2005})}\BibitemShut
  {NoStop}%
\bibitem [{\citenamefont {Benyamine}\ \emph {et~al.}(2017)\citenamefont
  {Benyamine}, \citenamefont {Aussillous},\ and\ \citenamefont
  {Dalloz-Dubrujeaud}}]{benyamine2017discharge}%
  \BibitemOpen
  \bibfield  {author} {\bibinfo {author} {\bibfnamefont {M.}~\bibnamefont
  {Benyamine}}, \bibinfo {author} {\bibfnamefont {P.}~\bibnamefont
  {Aussillous}},\ and\ \bibinfo {author} {\bibfnamefont {B.}~\bibnamefont
  {Dalloz-Dubrujeaud}},\ }\bibfield  {title} {\bibinfo {title} {Discharge flow
  of a granular media from a silo: effect of the packing fraction and of the
  hopper angle},\ }\href@noop {} {\bibfield  {journal} {\bibinfo  {journal}
  {EPJ Web of Conferences}\ }\textbf {\bibinfo {volume} {140}},\ \bibinfo
  {pages} {03043} (\bibinfo {year} {2017})}\BibitemShut {NoStop}%
\bibitem [{\citenamefont {Perge}\ \emph {et~al.}(2012)\citenamefont {Perge},
  \citenamefont {Aguirre}, \citenamefont {Gago}, \citenamefont {Pugnaloni},
  \citenamefont {Le~Tourneau},\ and\ \citenamefont
  {G{\'e}minard}}]{perge2012evolution}%
  \BibitemOpen
  \bibfield  {author} {\bibinfo {author} {\bibfnamefont {C.}~\bibnamefont
  {Perge}}, \bibinfo {author} {\bibfnamefont {M.~A.}\ \bibnamefont {Aguirre}},
  \bibinfo {author} {\bibfnamefont {P.~A.}\ \bibnamefont {Gago}}, \bibinfo
  {author} {\bibfnamefont {L.~A.}\ \bibnamefont {Pugnaloni}}, \bibinfo {author}
  {\bibfnamefont {D.}~\bibnamefont {Le~Tourneau}},\ and\ \bibinfo {author}
  {\bibfnamefont {J.-C.}\ \bibnamefont {G{\'e}minard}},\ }\bibfield  {title}
  {\bibinfo {title} {Evolution of pressure profiles during the discharge of a
  silo},\ }\href@noop {} {\bibfield  {journal} {\bibinfo  {journal} {Physical
  Review E—Statistical, Nonlinear, and Soft Matter Physics}\ }\textbf
  {\bibinfo {volume} {85}},\ \bibinfo {pages} {021303} (\bibinfo {year}
  {2012})}\BibitemShut {NoStop}%
\bibitem [{\citenamefont {Samadani}\ \emph {et~al.}(1999)\citenamefont
  {Samadani}, \citenamefont {Pradhan},\ and\ \citenamefont
  {Kudrolli}}]{samadani1999size}%
  \BibitemOpen
  \bibfield  {author} {\bibinfo {author} {\bibfnamefont {A.}~\bibnamefont
  {Samadani}}, \bibinfo {author} {\bibfnamefont {A.}~\bibnamefont {Pradhan}},\
  and\ \bibinfo {author} {\bibfnamefont {A.}~\bibnamefont {Kudrolli}},\
  }\bibfield  {title} {\bibinfo {title} {Size segregation of granular matter in
  silo discharges},\ }\href@noop {} {\bibfield  {journal} {\bibinfo  {journal}
  {Physical Review E}\ }\textbf {\bibinfo {volume} {60}},\ \bibinfo {pages}
  {7203} (\bibinfo {year} {1999})}\BibitemShut {NoStop}%
\bibitem [{\citenamefont {Chen}\ \emph {et~al.}(2021)\citenamefont {Chen},
  \citenamefont {Li}, \citenamefont {Xiu}, \citenamefont {Zivkovic},\ and\
  \citenamefont {Yang}}]{chen2021measurement}%
  \BibitemOpen
  \bibfield  {author} {\bibinfo {author} {\bibfnamefont {Q.}~\bibnamefont
  {Chen}}, \bibinfo {author} {\bibfnamefont {R.}~\bibnamefont {Li}}, \bibinfo
  {author} {\bibfnamefont {W.}~\bibnamefont {Xiu}}, \bibinfo {author}
  {\bibfnamefont {V.}~\bibnamefont {Zivkovic}},\ and\ \bibinfo {author}
  {\bibfnamefont {H.}~\bibnamefont {Yang}},\ }\bibfield  {title} {\bibinfo
  {title} {Measurement of granular temperature and velocity profile of granular
  flow in silos},\ }\href@noop {} {\bibfield  {journal} {\bibinfo  {journal}
  {Powder Technology}\ }\textbf {\bibinfo {volume} {392}},\ \bibinfo {pages}
  {123} (\bibinfo {year} {2021})}\BibitemShut {NoStop}%
\bibitem [{\citenamefont {Zuriguel}\ \emph {et~al.}(2019)\citenamefont
  {Zuriguel}, \citenamefont {Maza}, \citenamefont {Janda}, \citenamefont
  {Hidalgo},\ and\ \citenamefont {Garcimart{\'\i}n}}]{zuriguel2019velocity}%
  \BibitemOpen
  \bibfield  {author} {\bibinfo {author} {\bibfnamefont {I.}~\bibnamefont
  {Zuriguel}}, \bibinfo {author} {\bibfnamefont {D.}~\bibnamefont {Maza}},
  \bibinfo {author} {\bibfnamefont {A.}~\bibnamefont {Janda}}, \bibinfo
  {author} {\bibfnamefont {R.~C.}\ \bibnamefont {Hidalgo}},\ and\ \bibinfo
  {author} {\bibfnamefont {A.}~\bibnamefont {Garcimart{\'\i}n}},\ }\bibfield
  {title} {\bibinfo {title} {Velocity fluctuations inside two and three
  dimensional silos},\ }\href@noop {} {\bibfield  {journal} {\bibinfo
  {journal} {Granular Matter}\ }\textbf {\bibinfo {volume} {21}},\ \bibinfo
  {pages} {1} (\bibinfo {year} {2019})}\BibitemShut {NoStop}%
\bibitem [{\citenamefont {Walker}(1966)}]{walker1966approximate}%
  \BibitemOpen
  \bibfield  {author} {\bibinfo {author} {\bibfnamefont {D.}~\bibnamefont
  {Walker}},\ }\bibfield  {title} {\bibinfo {title} {An approximate theory for
  pressures and arching in hoppers},\ }\href@noop {} {\bibfield  {journal}
  {\bibinfo  {journal} {Chemical Engineering Science}\ }\textbf {\bibinfo
  {volume} {21}},\ \bibinfo {pages} {975} (\bibinfo {year} {1966})}\BibitemShut
  {NoStop}%
\end{thebibliography}%

\end{document}